\documentclass[twocolumn,showpacs,preprintnumbers,amsmath,amssymb,prb]{revtex4}

\usepackage{graphicx}
\usepackage{dcolumn}
\usepackage{bm}
\usepackage{amsmath}
\usepackage{placeins}
\usepackage{xcolor}
\newcommand{\yk}[1]{{\color{black} #1}}

\begin{document}


\title{Origins of conductance anomalies in a p-type GaAs quantum point contact}

\author{Y. Komijani\footnote{Now at Department of Physics and Astronomy, University of British Columbia, Vancouver, B.C., Canada V6T 1Z1}}
\email{komijani@phys.ethz.ch}
\affiliation{Solid State Physics Laboratory, ETH Zurich, 8093 Zurich, Switzerland}

\author{M. Csontos\footnote{Now at Department of Physics, Budapest University of Technology and Economics, 1111 Budapest, Hungary}
}
\affiliation{Solid State Physics Laboratory, ETH Zurich, 8093 Zurich, Switzerland}
\author{T. Ihn}
\author{K. Ensslin}
\affiliation{Solid State Physics Laboratory, ETH Zurich, 8093 Zurich, Switzerland}
\author{Y. Meir}
\affiliation{Physics Department, Ben Gurion University, Beer Sheva 84105, Israel}
\author{D. Reuter\footnote{Now at University Paderborn, Department Physik, Warburger Stra\ss e 100, 33098 Paderborn, Germany}}
\author{A. D. Wieck}
\affiliation{Angewandte Festk\"orperphysik, Ruhr-Universit\"at Bochum, 44780 Bochum, Germany}

\date{\today}

\begin{abstract}

Low temperature transport measurements on a p-GaAs quantum point contact are presented which reveal the presence of a conductance anomaly that is markedly different from the conventional `0.7 anomaly'. A lateral shift by asymmetric gating of the conducting channel is utilized to identify and separate different conductance anomalies of local and generic origins experimentally. While the more generic 0.7 anomaly is not directly affected by changing the gate configuration, a model is proposed which attributes the additional conductance features to a gate-dependent coupling of the propagating states to localized states emerging due to a nearby potential imperfection. Finite bias conductivity measurements reveal the interplay between the two anomalies consistently with a two-impurity Kondo model.

\end{abstract}

\maketitle


\section{Introduction}
Since its discovery \cite{Wees88,Wharam88} conductance quantization in ballistic quantum point contacts (QPCs) has become one of the hallmarks of mesoscopic physics. This quantization has a single-particle origin and is due to lateral confinement and cancellation of group velocity and density of states in one dimension. In addition to the conductance plateaus at integer multiples of $2e^2/h$, QPCs often exhibit features which require an explanation beyond the single-particle picture provided by the Landauer theory, the most prominent one being the 0.7-anomaly.~\cite{Thomas96} The latter is usually referred to as an extra anomalous plateau at the conductance value of about 0.7$(\times 2e^2/h)$. Despite its name, it is not the precise conductance value of this plateau but rather a set of qualitative features that is usually associated with the anomaly.~\cite{Micolich11} The feature evolves smoothly into the spin-resolved $e^2/h$ plateau at high in-plane magnetic fields \cite{Thomas96,Wieck96} revealing its spin-related nature. Moreover, while the conductance quantization gets weaker by increasing the temperature due to thermal smearing, the anomalous plateau typically gets stronger and drops in conductance.~\cite{Thomas96, Kristensen00, Cronenwett02,Komijani10} Furthermore, close to pinch-off QPCs usually exhibit non-linearities and especially a peak around zero bias in their $dI/dV$ known as the zero bias anomaly (ZBA).~\cite{Kristensen00} The empirical correlation between the 0.7-anomaly and the ZBA has suggested a connection between the two \cite{Cronenwett02} although this correlation and conclusions thereof have been debated.~\cite{Sfigakis08,Sarkozy09,Chen09,Ren10,Liu11}

Possible explanations to date have been based on many-body phenomena including spontaneous spin polarization,~\cite{Berggren02,Reilly02,Reilly05,Rokhinson06} separation of singlet and triplet channels \cite{Rejec00} or spin and charge channels,~\cite{Matveev04} electron-electron interaction \cite{Lunde09} and Coulomb repulsion arising from a quasi-localized state forming in the QPC.~\cite{Cronenwett02, Meir02, Meir08} The latter includes Kondo screening of an unpaired spin and a restoration of the conductance to $2e^2/h$ at low temperature due to the Kondo effect, consistent with the observation of the ZBA.

In spite of extensive studies, a complete understanding of the conductance anomalies in QPCs is still {\yk missing}. One technical reason is that some of the features associated with the 0.7-anomaly can also arise from extrinsic sources, like impurities or potential imperfections, and may overshadow the features due to the generic 0.7-anomaly.~\cite{McEuen90,Wees91,Nockel94,Sfigakis08} These imperfections can even arise during the fabrication process and are thus not totally excluded by using a high mobility heterostructure. Thus the experimental separation of such impurity related contributions from the generic features is rather crucial. Such a differentiation is actually not trivial and may require exploring the whole experimental parameter space. Here we demonstrate the co-existence of the 0.7-anomaly and a signal with an extrinsic origin, show how to separate the two, and discuss what that extrinsic origin may be.

We study transport properties of a QPC implemented in p-type GaAs and demonstrate that it exhibits conductance anomalies which are very similar to a classic 0.7-feature. After a thorough analysis of the data and a comparison with a number of simple models we conclude that one of the anomalies observed here is \emph{different} than the conventional 0.7-anomaly and is caused by potential fluctuations. 
The presented data correspond to the rare situation where such a detailed understanding and thereby the separation of the anomalies arising from different origins is possible. 
Furthermore, the results presented here suggest both coexistence and the interplay between an impurity resonance and the 0.7-anomaly.

The more pronounced carrier-carrier interactions in p-type QPCs compared to their n-type counterparts make them especially suitable for investigating many-body effects such as the 0.7-anomaly.~\cite{Komijani10} The valence band holes are spin-3/2 carriers yielding interesting consequences, e.g. the recent observation of non-conventional Kondo physics in these systems.~\cite{Klochan11} Moreover, the holes' strong spin-orbit interaction leads to the peculiar property that their $g$-factor is influenced by the lateral confinement.~\cite{Danneau08,Chen10} This confinement anisotropy opens a possibility to test a conjectured connection between the conductance anomalies and the spin subbands.~\cite{Danneau06,Hamilton08} 

\yk {Recently we have shown by studying a number of p-type QPCs~\cite{Komijani10} that the 0.7-anomaly in these samples is transformed to a narrow conductance peak accompanied by a diamond-like suppression of the differential conductance under the application of a magnetic field perpendicular to the transport plane. While this effect has been reproduced in many different QPCs and is believed to be a generic effect, the features reported in this article are special to this particular sample (but robust against thermal cycling).}
\section{Sample and setup}

\begin{figure}
\includegraphics[width=0.45\textwidth]{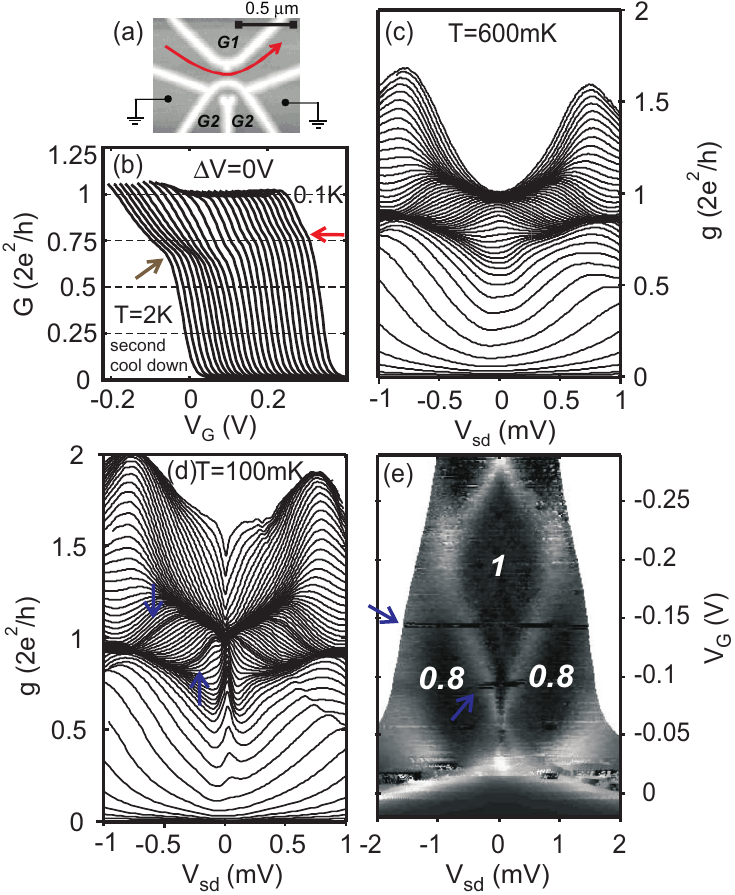}
\caption{\label{fig:Fig1}\small{(color online) (a) AFM micrograph of the sample. The transport through the point contact channel is controlled by the application of voltages on two surrounding conducting areas G1 and G2, while all the other gates are grounded. (b) Temperature-dependence of the conductance anomaly from 0.1~K (right) to 2~K (left). \yk{The conductance value of the anomaly decreases with increasing temperature}. This
 measurement was performed at $\Delta V$~=~0~V (see text for details). Curves are shifted horizontally for clarity. (c) and (d) Non-linear differential conductance $g=dI/dV_{sd}$ as a function of the bias voltage for different gate voltages at 600~mK and 100~mK. The first plateau has an upward slope with $V_{sd}$ in both cases. In (d) this upward slope is more extreme and linear and a peak at zero bias is also evident. The $dI/dV_{sd}$ curves show a clear accumulation, called bunching here, around two gate voltages shown with blue arrows. (e) Transconductance ${d^2I}/{dV_{sd}dV_{g}}$ (arbitrary units) as a function of source-drain bias and gate voltage obtained by numerical differentiation from the data presented in Fig 1(d) over larger gate and source-drain voltage ranges. The nonlinear conductances are indicated in units of $2e^2/h$. The bunchings marked with blue arrows are also visible in this plot. 1(c-d) were measured at $V_G=V_{G1}=V_{G2}$ ($\Delta V$ was changed by 64~mV during this gate sweep, which is small compared to the $\Delta V$ variation of 0.8\,V in Fig.\,\ref{fig:Fig2}(a) and 0.3\,V in the rest of the paper).}}
\end{figure}

Fig.\,\ref{fig:Fig1}(a) shows the AFM micrograph of the sample which was patterned in the two-dimensional hole gas (2DHG) by local anodic oxidation lithography.~\cite{Held98,Rokhinson02} The bright oxide lines displayed in Fig.\,\ref{fig:Fig1}(a) locally deplete the 2DHG situated 45~nm below the surface separating the 2DHG-plane into laterally disconnected regions which are individually connected to metallic leads. The host heterostructure consists of a C-doped GaAs/AlGaAs heterostructure grown along the (100) plane.~\cite{Wieck00} Prior to sample fabrication the quality of the 2DHG was characterized by standard magnetotransport measurements at 4.2~K and the hole density $n=4\times 10^{11}$~cm$^{-2}$, and mobility $\mu$ = 120'000 cm$^2$/Vs were obtained.

The QPC studied here has a lithographical width of 150~nm whose geometric confinement is asymmetric with respect to its transport axis, as displayed in Fig.~\ref{fig:Fig1}(a), and can be tuned by the G1 and G2 in-plane gates. The linear combination $V_{g}=\alpha_{1}V_{G1}+\alpha_{2}V_{G2}$ tunes the confining potential in a symmetric fashion, where $-\alpha_{1}/\alpha_{2}$ is the slope of the lines of equal linear conductance in the $V_{G1}$\,--\,$V_{G2}$ plane. Accordingly, the asymmetric gate voltage combination $\Delta V=\alpha_{2}V_{G1}-\alpha_{1}V_{G2}$ leads to a transverse electric field resulting in a lateral shift of the channel while the \yk{density} is kept constant.

Transport experiments were carried out between $T=$~2~K and the 100~mK base temperature of a $^{3}$He/$^{4}$He dilution refrigerator at magnetic fields of up to $B$~=~13~T applied both perpendicular to the plane and in the plane (perpendicular to the current) of the QPC. Measurements of the finite-bias differential conductance $g=dI(V_{sd},V_{g})/dV_{sd}$ were carried out by the simultaneous symmetric application of an ac excitation between source and drain with an amplitude of 20 $\mu$V at 31 Hz lock-in frequency, and a dc offset $V_{\rm bias}$ of up to 6 mV. Four-terminal lock-in measurements of the linear conductance ($G={g}(V_{sd}=0)$) were performed at a frequency of 31~Hz. The voltage drop $V_{sd}$ across the QPC was measured between two independent leads not used for the application of the bias voltage.

Most of the data presented here were acquired in the first cool down. The sample was cooled down a second time in order to extend the range of temperature-dependent measurements and to change the orientation of the sample with respect to the magnetic field. Figures showing measurements of the second cool down are labelled accordingly. The general results discussed in this article have been qualitatively reproduced in three further cool downs as well.

\section{Results and discussion}

\subsection{Temperature dependence}
Fig.\,\ref{fig:Fig1}(b) shows the linear conductance of the QPC as a function of the (symmetric) voltage applied to the surrounding gates $V_{g}$ measured at temperatures ranging from 0.1~K to 2~K. The conductance is tuned between the pinch-off regime at positive gate voltage and the first plateau as the gate voltage is lowered. Due to the small subband energy spacing of holes no conductance quantization is visible at $T=$~2~K but there is a clear kink in the linear conductance marked with the black (brown online) arrow. At lower temperatures, however, this kink disappears and a plateau appears at $G=2e^2/h$. The full length of the first plateau cannot be seen in the figure as the accessible gate voltage range is limited by leakage currents. The rightmost curve taken at $T=$~0.1~K also exhibits a faint kink which is marked with the bright (red online) arrow. The detailed study of these two anomalies on the rise of the first plateau is the main topic of this paper. They resemble to the standard 0.7-anomaly in QPCs. Nevertheless, in the subsequent sections we demonstrate that one of them arises due to an extrinsic impurity. In the rest of the paper we focus on the anomaly at $T=$ 0.1\,K and its evolution within the range of $T<$ 1\,K and only discuss the $T>$ 1\,K case at the end of the paper.

\subsection{Finite bias spectroscopy}
The subband structure and the density of states in the QPC can be investigated by measuring $g=dI/dV_{sd}$ as a function of the applied bias voltage. The non-linear differential conductance ${g}$ of the QPC is shown in Fig.\,\ref{fig:Fig1}{(c)-(d)} at $T=$~600 and 100~mK, respectively. Each curve in this figure corresponds to a specific gate voltage $V_{G}$ and the plateaus in $G$(${V}_{{g}}$) appear as the accumulation of the individual $g(V_{sd},V_g)$ traces corresponding to different gate voltages.
Figures \ref{fig:Fig1}{(c)-(d) both show that the first plateau shifts to higher conductances with increasing bias. This upward shift which is more extreme and becomes a linear bias dependence at base temperature (Fig.\,\ref{fig:Fig1}(d)) \yk{is not yet understood}. The difference between these two measurements points to the emergence of an asymmetric zero bias anomaly (ZBA) at the base temperature.  Numerical differentiation of the $dI/dV_{sd}$ data shown in Fig.\,\ref{fig:Fig1}(d) with respect to the gate voltage results in the transconductance plotted in a gray-scale color map in Fig.\,\ref{fig:Fig1}(e). The transconductance displays the subband structure of the point contact confirming ballistic transport through the QPC. Two charge re-arrangements marked with blue arrows, visible in this figure are also seen in the differential conductance of Fig.\,\ref{fig:Fig1}(d) (again marked with blue arrows) as the accumulation of $dI/dV_{sd}$ traces of nearby gate voltages. The discontinuities in the differential conductance at the gate voltage corresponding to these two points happen presumably because of the (dis-)charging of a nearby impurity situated either in the doping layer or within the oxide lines, acting on the QPC channel as an additional gate. A similar mechanism is conventionally used for charge detection in quantum dots. These re-arrangements happen on the first plateau and do not affect the conductance anomaly discussed before.
\subsection{Lateral shift of the QPC channel}
Employing an asymmetric combination of the voltages applied to the two gates enables the lateral shift of the QPC channel as discussed before. This lateral shift is estimated to be 8-10 nm/V within the range of $\Delta V=\pm$~300~mV.~\cite{Heinzel00} Considering an impurity Bohr radius of the order of 1-2~nm this covers a sufficiently large region to reveal possible impurity resonance effects. Figure \ref{fig:Fig2}(a) shows the linear conductance of the QPC as a function of the symmetric gate voltage $V_{g}$ at different gate asymmetries $\Delta V$. The symmetric configuration ($\Delta V=$0) is marked with a thick curve. The magnitude of the anomalous feature is strongly modified by the gate asymmetry, disappearing at negative $\Delta V$ values and transforming to a small plateau for $\Delta V>$~0.1~V, although its conductance is not affected. 
The observed strong dependence on the lateral field implies that the anomalous feature is most probably due to impurities and/or potential imperfections in the channel. 

Fig.\,\ref{fig:Fig2}(b) shows the temperature-dependence of the linear conductance at $\Delta V=$~+113~mV for temperatures ranging from $T=$~100 to 600~mK. At this positive gate asymmetry the anomalous feature is a clear plateau whose conductance drops with increasing temperature. A similar temperature dependence is usually observed for the 0.7-feature in quantum point contacts and is often \yk{considered to be connected} with the amplitude of the ZBA.~\cite{Cronenwett02,Komijani10} Comparison of the ZBA at two different temperatures of $T=$~100~mK and $T=$~600~mK in Fig.\,\ref{fig:Fig1}(c,d) shows that the amplitude of the ZBA is indeed correlated with the temperature dependence of the anomalous feature that we observe in the linear conductance.

\begin{figure}
\includegraphics[width=0.45\textwidth]{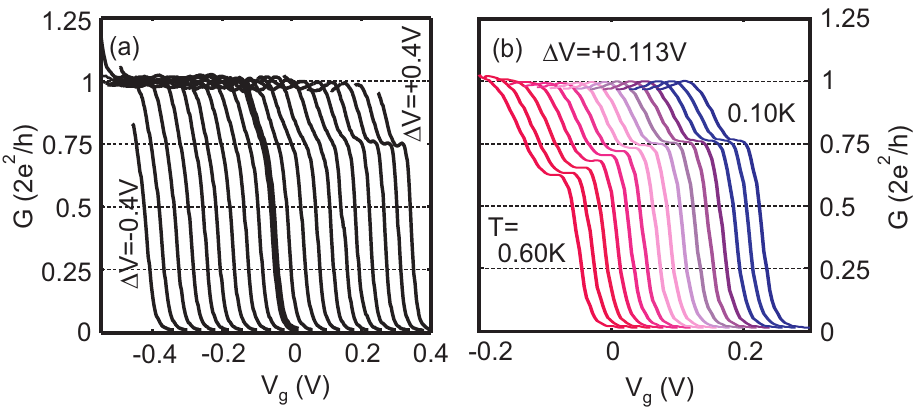}
\caption{\label{fig:Fig2}\small{(color online) (a) Linear conductance $G$ as a function of $V_{g}$ for different gate asymmetries ranging from $\Delta V=-0.4$~V to +0.4~V at T=100~mK. The symmetric configuration ($\Delta V=$~0) is marked with a thick line. Due to the break down across the oxide lines the gate voltage range is limited at extreme asymmetries. The conductance anomaly gets stronger for positive asymmetries (when the channel is shifted towards the gate $G2$), transforming to a plateau at $G\sim$0.75~$\times 2e^2/h$ for $\Delta V\sim$~100~mV and disappears at negative asymmetries. (b) Temperature-dependence of the linear conductance $G$ vs.~$V_{g}$ at $\Delta V=$~+113~mV in the range of 100~mK (right) to 600~mK (left). The conductance of the anomalous plateau decreases with increasing temperature. The curves are shifted horizontally for clarity.}}
\end{figure}
\subsection{Magnetic field dependence}
In order to gain more insight about the anomalous feature, the effect of in-plane and perpendicular magnetic fields on the conductance anomaly is reported here. Under an in-plane magnetic field a typical 0.7-feature evolves continuously into the spin-split plateau at 0.5$\times 2e^2/h$. The application of an in-plane magnetic field perpendicular to the current direction (Fig.\,\ref{fig:Fig3}(a)) also results in a drop in the conductance of the anomalous feature. However instead of evolving to the half-plateau, it crosses the half-plateau and saturates around 0.4$\times 2e^2/h$ at sufficiently large field.

It must be noted that due to heavy hole (HH) and light hole (LH) mixing, there is a considerable confinement anisotropy between $g$-factors in the two in-plane directions.~\cite{Danneau08,Chen10} At the particular orientation between the current and the in-plane magnetic field directions considered here the Zeeman effect is expected to be strongly suppressed.~\cite{Chen10}

\begin{figure}
\includegraphics[width=0.45\textwidth]{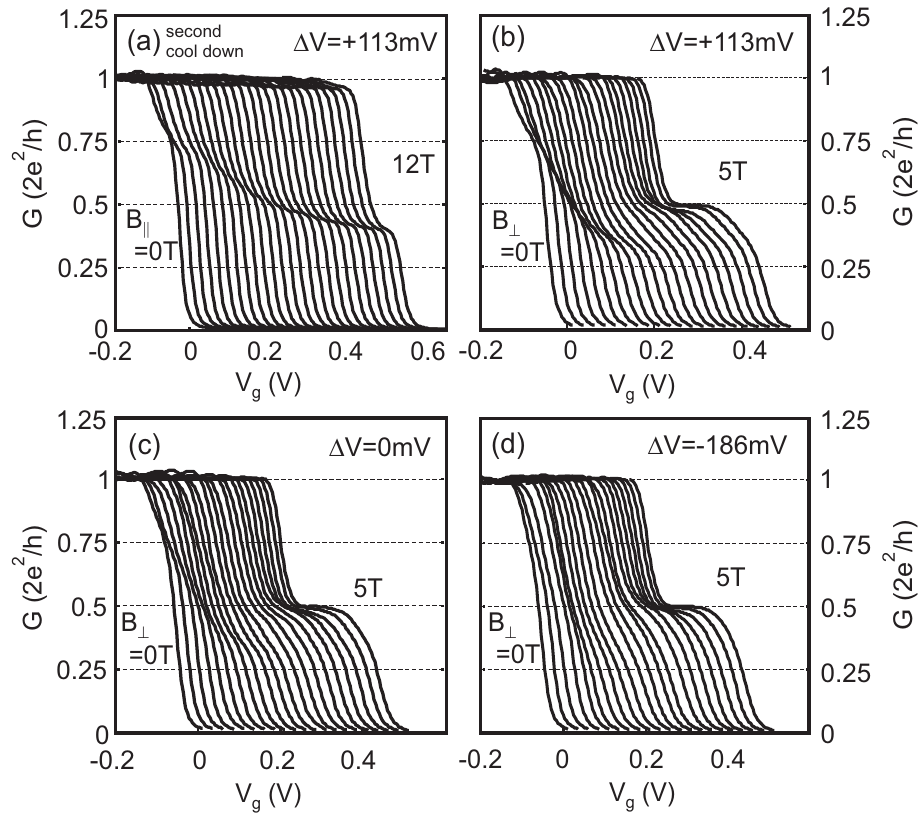}
\caption{\label{fig:Fig3}\small{(a) Linear conductance $G$ as a function of $V_{g}$ at different in-plane magnetic fields ranging from 0~T to 12~T at the gate asymmetry of $\Delta V$=+113~mV at T=100~mK. Note that the anomalous feature moves to lower conductances and even below 0.5$\times 2e^2$/h as the field is increased. A constant series resistance attributed to the leads was subtracted from the raw data. (b)-(d) The effect of a magnetic field perpendicular to the 2DHG plane on linear conductance $G$ vs.~$V_{g}$ at three different gate asymmetries of $\Delta V$~=+113, 0 and -186 mV. Each plot shows the effect of perpendicular magnetic fields up to 5~T. The same trend in the anomalous feature as shown in (a) is observed for (b) and (c). In all three cases, a seemingly independent half-plateau is formed at 0.5$\times 2e^2/h$ for $B_{\perp}>$~3~T. In case of perpendicular magnetic fields a $B_{\perp}$-dependent but gate-independent series resistance is subtracted from the raw resistance data which was checked to be due to SdH oscillations in the leads. The curves are shifted horizontally for clarity.}}
\end{figure}

Figure \ref{fig:Fig3}(b) shows the effect of the perpendicular magnetic field on the conductance anomaly at $\Delta V=$~+113~mV. A field-dependent series resistance is subtracted from the measured data which was confirmed to be consistent with the Shubnikov-de Haas (SdH) oscillations in the leads.~\cite{Komijani10} As the field increases, the conductance of the anomalous feature gradually decreases, crossing 0.5$\times 2e^2/h$ and saturating around 0.4$\times 2e^2/h$, very similar to the in-plane magnetic field case. At $B_{\perp}>$~3~T a separate and seemingly independent plateau appears at 0.5$(\times 2e^2/h)$. We have paid extra attention to make sure that the precise values of the large-field conductance of the anomalous plateau do not come from our resistance subtraction method. The maximum subtracted resistance is limited by the minimum resistance of each curve and is linked to our precision in the estimation of the first plateau and the half plateau. Figures \ref{fig:Fig3}(c)-(d) show similar experimental results concerning the $B_{\perp}$-dependence at gate asymmetries of $\Delta V=$~0 and -186~mV. The evolution of the spin-split subband is similar in these plots to the case of $\Delta V=$~+113~mV (Fig.\,\ref{fig:Fig3}(b)) but compared to the latter the anomalous feature gets weaker as the channel is shifted towards gate G1.

The difference between the data shown in Fig.\,\ref{fig:Fig3}(a) and (b) arises due to the lower value of the in-plane $g$-factor ($g_{\parallel}<g_{\perp}$), however, the fact that the anomalous feature crosses the conductance of the half-plateau is clearly visible in both cases. The features present in an in-plane field of $B_{\parallel}\sim$~12~T are already developed at the perpendicular field of $B_{\perp}\sim$~2.5~T. From this we estimate a ratio of $\approx$~5 for the two corresponding $g$-factors. Since the $g$-factor of 1D subbands is expected to be zero for in-plane magnetic fields perpendicular to the QPC axis, we speculate that our non-zero value is the consequence of a possible misalignment between the field and the QPC axis and/or the crystallographic directions.

\FloatBarrier
\subsection{Zero bias anomaly} \label{sec:zba}
It has been suggested~\cite{Cronenwett02, Meir02, Meir08} that the ZBA can be caused by the presence of a peak in the \yk{local density of states} of the QPC due to the Kondo effect, very similar to what happens with an unpaired spin in the strong coupling regime of a quantum dot. However, a zero bias peak can also have non-Kondo origins \cite{Ghosh05,Sedrakyan07,Chen09,Sarkozy09,Ren10} and, to this date, there is no consensus about the source of the ZBA. To shed more light on possible origins of the ZBA observed here, we study the effect of $\Delta V$ on the ZBA in this section. As expected, the shape of the ZBA also depends on the lateral shift of the channel. Figure \ref{fig:Fig4}(a)-(c) show the low bias part of the differential conductance at various values of $\Delta V$. While a clean and isolated zero bias peak is observed between pinch-off and the first plateau at negative $\Delta V$, it becomes slightly asymmetric at $\Delta V=0$. At the same time several $dI/dV_{sd}$ curves corresponding to different gate voltages get closer in conductance. At a positive asymmetry of $\Delta V$~=+113~mV the ZBA is split into two peaks and a bunch of $dI/dV_{sd}$ curves are accumulated to form a plateau-like structure that follows the shape of the zero bias peak near $G\sim 0.8\times 2e^2/h$. This plateau is the finite-bias signature of the anomaly we observed in the linear conductance before. Note that the ZBA is more symmetric above this plateau. The zero-field asymmetry of the ZBA at $\Delta V$~=-186~mV can be understood within a Kondo picture in terms of a coupling asymmetry of the impurity to the leads.~\cite{Krawiec02}

\begin{figure}
\includegraphics[width=0.45\textwidth]{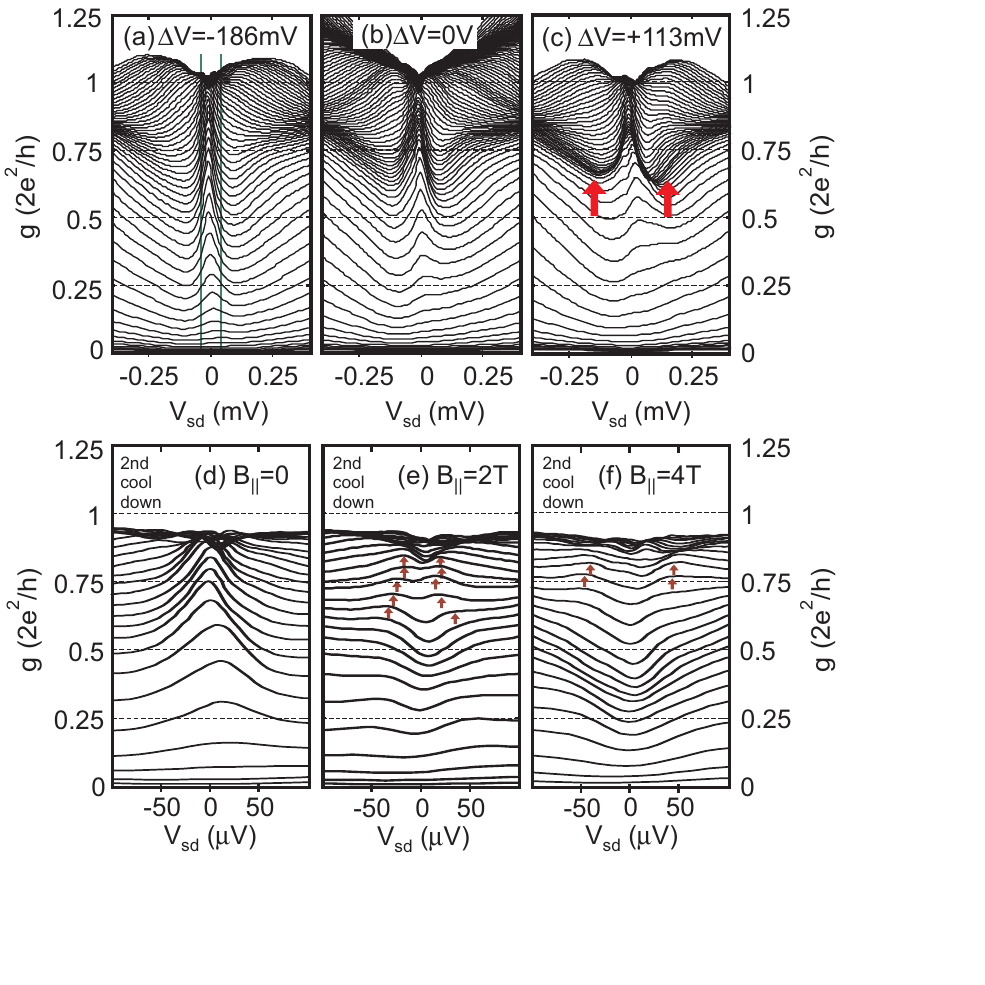}
\caption{\label{fig:Fig4}\small{(color online) (a)-(c) Differential conductance $g$ vs.~$V_{sd}$ and $V_g$ exhibiting a ZBA at three different gate asymmetries $\Delta V$. At $\Delta V$~=~-186~mV the ZBA is single-peaked and only slightly asymmetric. At zero and positive $\Delta V$ values the peak becomes more asymmetric and splits. The arrows point to the accumulation of the individual $dI/dV_{sd}$ curves at $\Delta V$~=~+113~mV. A zoom into the low bias regime at $\Delta V$~=~-186~mV (marked by the green vertical lines in (a)) is shown in (d)-(f) for different in-plane magnetic fields. The ZBA splits at $B_{\parallel}$=~2~T and the splitting grows further at 4~T. All data are taken at $T=$~100~mK.}}
\end{figure}

In order to see whether the ZBA observed here is related to the Kondo effect, we investigate the effect of an in-plane magnetic field on the splitting of the peak as shown in Fig.\,\ref{fig:Fig4}(d)-(f). Here a negative $\Delta V$ is chosen for which the ZBA is symmetric. 
The single peak at zero field splits at $B_{\parallel}$=~2~T and the splitting grows further at $B_{\parallel}$=~4~T. The peaks are marked by small brown arrows. These splittings can be compared to the $g$-factor. In a Kondo system such a splitting is expected to be twice the Zeeman energy \cite{Meir93} and therefore the results would be consistent with a Kondo picture if $g_{\parallel}\sim 0.2$. 
This value is reasonable in a QPC where the current is not perfectly perpendicular to the in-plane magnetic field and is consistent with the values reported in the literature.~\cite{Chen10}


\subsection{Two models of an impurity}

In this section we compare the experimentally observed signatures of the anomaly discussed so far with two simplified models for the impurity. The first is a single-particle model employed in the framework of the Landauer theory, where transport properties can be calculated from a transmission function. Here we model a potential imperfection by assuming the transmission to be a Lorentzian resonance superimposed on top of the transmission of a saddle-point potential~\cite{Glazman89,Buttiker90,MartinMoreno92} as shown in Fig.\,\ref{fig:Fig5}(a). The temperature dependence shown in Fig.\,\ref{fig:Fig5}(b) implies that such transmission resonances disappear quickly by increasing the temperature. It is interesting though that the resonances lead to a zero bias peak in the differential conductance as shown in Fig.\,\ref{fig:Fig5}(c) that also disappears by increasing the temperature (not shown). Differential conductance traces shown in this figure are calculated assuming a symmetric bias voltage applied to the QPC. The assumption that the energy-dependence of the transmission is the same for all the subbands (except for a relative shift by the subband spacing) results in the appearance of the zero bias peak at all the three rises of the differential conductance shown in the figure. Taking into account screening effects would suppress the higher zero bias peaks. Note that there are also curves with split zero-bias peaks in the zero-field differential conductance.

Application of an in-plane magnetic field splits the subbands and since we assumed equal transmission for spin-split subbands, the resonance appears in both subbands as shown in Fig.\,\ref{fig:Fig5}(d). The finite bias spectroscopy at finite magnetic field shows similar pattern as the zero-field one, but here the split zero-bias peaks are even more evident. Therefore a zero bias peak can appear in $g$ due to a transmission resonance, get suppressed by increasing the temperature and may even split at certain gate voltages with or without the application of an in-plane magnetic field. A qualitative discussion on the origin of the different features arising in this model is presented in the appendix.

\begin{figure}
\includegraphics[width=0.45\textwidth]{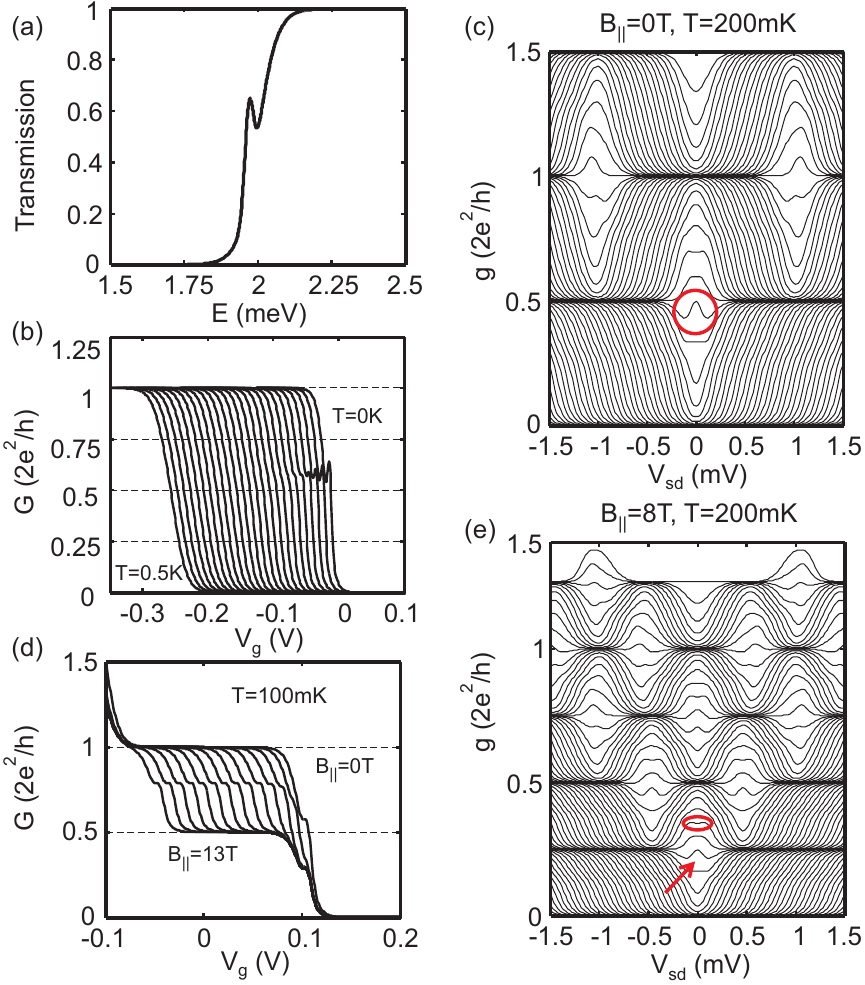}
\caption{\label{fig:Fig5}\small{(color online) Characteristics of a model QPC with a resonance in the transmission. (a) Transmission vs.~energy is assumed to have a Lorentzian resonance at the Fermi energy superimposed on the transmission of a saddle-point potential. (b) Calculated linear conductance vs.~$V_{g}$ for equally-spaced temperatures in steps of 20~mK. The resonance disappears quickly by increasing the temperature (shifted horizontally for clarity). (c) Non-linear differential conductance calculated from the transmission shows that a peak appears at zero bias. (d) Effect of an in-plane magnetic field on the linear conductance. (e) Effect of an in-plane magnetic field on the differential conductance.}}
\end{figure}

Another type of imperfection that we consider is the capacitive coupling between the QPC channel and a potential well or an isolated state that can get occupied at a certain gate voltage. The QPC has a saddle-point transmission but the potential of the nano-structure shown in Fig.\,\ref{fig:Fig6}(a) has a minimum with a certain state that can get filled by an electron/hole from either the channel or the leads. We assume that this state is weakly coupled to its environment and its occupancy is given by a Fermi-Dirac distribution. Upon filling the impurity state at a certain gate voltage the conductance trace of the QPC is shifted along gate voltage axis proportionally to their capacitive coupling and a feature appears in the linear conductance as shown in Fig.\,\ref{fig:Fig6}(b). Increasing the temperature leads to two effects: first, the conductance of the QPC decreases at this gate voltage and therefore the conductance value of the feature also gradually decreases, very similarly to the `standard' 0.7 anomaly. Second, it extends the width of the feature by broadening the occupation profile of the impurity in gate voltage. This occupation profile is given by the depth of the potential well and the temperature. For temperatures above the ionization energy of the impurity, not considered here, the anomaly is expected to vanish. Figure \ref{fig:Fig6}(c) shows differential conductance data which look very similar to those of an ideal QPC. The only difference is the bunching of several $dI/dV_{sd}$ curves at the gate voltages corresponding to the filling of the impurity (see red arrows). This feature is similar to the results presented in Fig.\,\ref{fig:Fig1}(d). Note that in contrast to the experimental data, the model assumed here does not produce any zero bias peak.

The most interesting feature of this model is that the application of a Zeeman field, shown in Fig.\,\ref{fig:Fig6}(d), results in a shift of the anomaly very similarly to the data shown in Fig.\,\ref{fig:Fig3}(a). A finite Zeeman splitting 
(with a $g$-factor of 2) is assumed for the subband but not for the localized charge state. As the spin-degenerate subbands split gradually only one subband is affected by the impurity filling. Starting from 0.75$\times 2e^2/h$ the conductance anomaly crosses 0.5$\times 2e^2/h$, saturates at 0.4$\times 2e^2/h$ and a seemingly independent half-plateau appears at higher fields, all very consistent with the data presented in Fig.\,\ref{fig:Fig3}(b)-(d). The comparison of the model with the perpendicular magnetic field experiment is based on the assumption that the effect of the perpendicular magnetic field on the QPC within the range $B_{\perp}<$ 5\,T applied in the experiment and after subtracting the $B_{\perp}$-dependent series contact resistance is mainly a Zeeman splitting and the orbital shift of the subbands and the impurity level. The important parameter is their relative shift which is included in the non-zero g-factor of 1D subbands.

\begin{figure}
\includegraphics[width=0.45\textwidth]{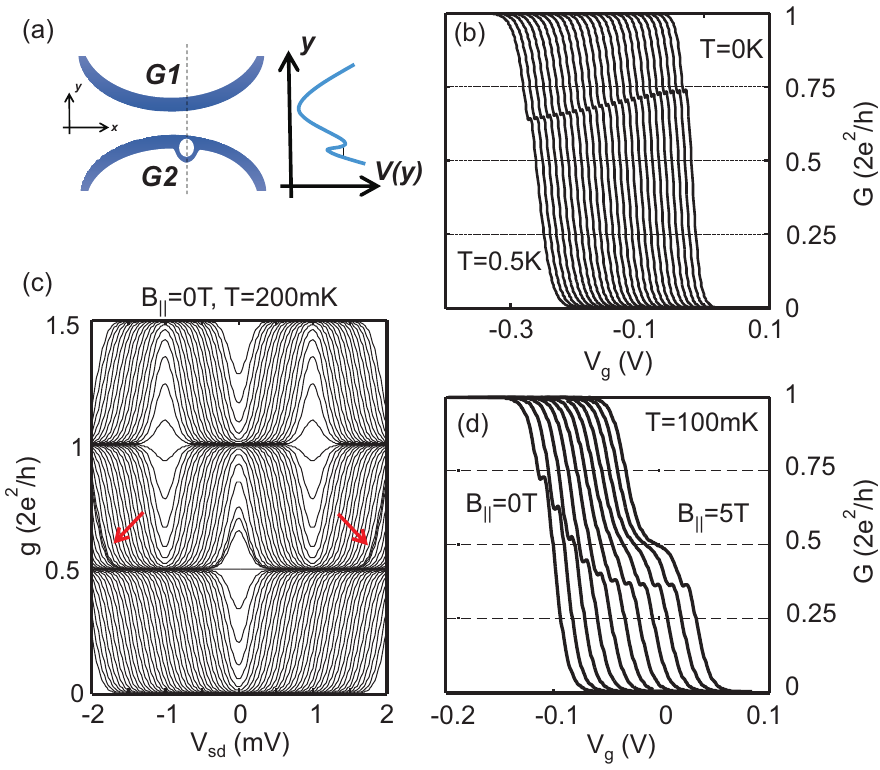}
\caption{(color online) Characteristics of a model QPC with a capacitive coupling to a nearby chargeable impurity. (a) Schematic of the model with an assumed potential profile along the dashed line (b) Calculated linear conductance vs.~$V_{g}$ for equally-spaced temperatures in steps of 20~mK. The anomaly in the conductance arises due to the charging of the impurity (shifted horizontally for clarity). (c) Non-linear differential conductance calculated from the transmission. The red arrows point to the bunchings of several curves at the gate voltage corresponding to the charging of the impurity. (d) Effect of an in-plane magnetic field on the linear conductance (shifted horizontally for clarity).\label{fig:Fig6}}
\end{figure}
	
\subsection{Possible explanation for T~$<$~1~K}

After reviewing these basic impurity effects on the characteristics of a QPC, we suggest the following picture. Since the anomalous feature is found to appear/disappear by the lateral shift of the channel, it is plausible to assume that some impurity or potential imperfections of the sample is involved. The model of a QPC with a nearby charge impurity fits quite well to the temperature-dependent and magnetic field-dependent data. The effect of the lateral shift of the channel can be accounted for by a $\Delta V$-dependence of the capacitive coupling of the impurity to the QPC channel. Based on the measurement shown in Fig.\,\ref{fig:Fig2}(a), we conclude that an impurity might be located near gate G2. By applying a positive (negative) $\Delta V$ the channel is pushed towards (pulled back from) the impurity. The only experimental feature that defies this simple model is the presence of a strong ZBA within a large gate voltage range. Considering the observed features of the ZBA, we assume that it has a different origin than the anomalous conductance feature discussed here.

\subsection{Local origin of the ZBA}

Adapting the model of a chargeable impurity, Fig.\,\ref{fig:Fig4} suggests that as the channel is pushed towards the impurity, the ZBA tends to get asymmetric and eventually splits into two peaks. Further details of the data are presented in Fig.\,\ref{fig:Fig7} where the effect of a magnetic field perpendicular to the plane of the 2DHG on the ZBA (after subtracting a $B_{\perp}$-dependent series resistance) is also shown in addition to the $\Delta V$-dependence. The top row with $\Delta V$~=~-286~mV (channel far from the impurity) shows a symmetric ZBA that starts to split around $B_{\perp}$~=~0.4~T. As the channel is pushed towards the impurity (more positive $\Delta V$) some of the $dI/dV_{sd}$ curves are bunched together. This corresponds to the onset of the occupation of the impurity in the model of a charge-able impurity discussed in Fig.\,\ref{fig:Fig6}. At conductances \emph{below} these bunchings (more positive gate voltage) the state is above the electrochemical potential of the leads and it is \emph{empty} while at conductances \emph{above} these bunchings the state is \emph{filled}. When the channel is close to the impurity ($\Delta V$~=~+300~mV) the ZBA is split even at $B=0$. It is interesting that this zero-field splitting of the ZBA happens only when the impurity state is empty (conductances below the bunching). Above that the ZBA is symmetric and single peaked. By applying a magnetic field the zero-field splitting of the empty-state ZBA grows even further. At $B_{\perp}\sim$~0.3~T, the ZBA of the filled-state also starts to split similarly to the behavior displayed in the first row of Fig.\,\ref{fig:Fig7}.

Several publications have associated the ZBA with the leads and not with the QPC.~\cite{Ghosh05,Sedrakyan07} Notwithstanding the questions about the origin of the ZBA, the zero-field splitting reported here supports a \emph{local origin of the} ZBA with a position located inside or in the close vicinity of the QPC.

\begin{figure}
\includegraphics[width=0.45\textwidth]{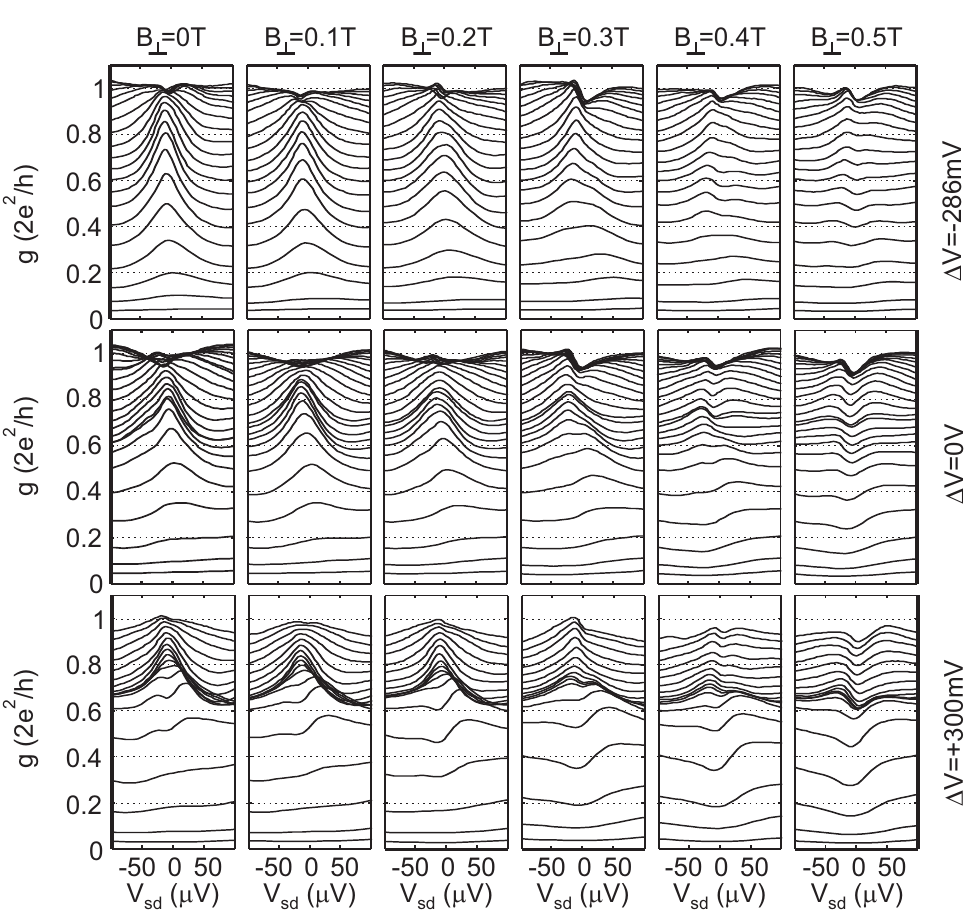}
\caption{\label{fig:Fig7} Evolution of the ZBA at $T=$~100~mK with perpendicular magnetic field $B_{\perp}$ and gate asymmetry $\Delta V$. A field-dependent series resistance has been subtracted to keep the first plateau at $2e^2/h$.}
\end{figure}

\subsection{Two-impurity Kondo system and crossover to higher temperatures}

If the ZBA is due to the Kondo effect, as suggested extensively in the literature \cite{Cronenwett02,Meir02,Meir08} and the splitting with in-plane magnetic field indicates, it is independent of the impurity. The Kondo theories of the 0.7 anomaly predict that the ZBA is a signature of the Kondo screening of an unpaired spin occupying a quasi-bound state forming in the QPC.~\cite{Meir08} The system investigated here is then very similar to the theoretically extensively investigated settings of the Kondo effect in a parallel double-dot \cite{Silva06,Silva08b,Fang10,Wang11} or in a single dot side-coupled to another dot.~\cite{Chung08,Zitko10} In these models the ZBA splits as a result of the hybridization between the two dots in qualitative agreement with our observations, supporting the idea that in addition to the capacitive coupling between the impurity and the channel, a $\Delta V$-dependent \emph{tunnel coupling} also exists between the two leading to a \emph{two-impurity Kondo behavior}. The relevant parameter in such a system is the coupling between the two spins (denoted here by $I$). Below a critical value of the coupling ($I<I^*$) the two spins are screened independently. For $I>I^*$ there is a quantum phase transition to an anti-ferromagnetic phase in which the two spins form a singlet.~\cite{Bork11} Assuming a Kondo origin of the ZBA, the presented data suggest that the inter-spin coupling $I$ is controlled by $\Delta V$ in our system. Additionally, the occupancy of the spurious impurity is controlled by the gate voltage $V_g$ and is expected to be equal to one in the gate voltage range below the bunching shown in Fig.\,7 explaining the observation of the ZBA for single-occupancy of the spurious impurity and sufficiently large $I$ ($\Delta V\sim$0.3\,V).

Note that we measure $dI/dV$ through the generic Kondo impurity which is side-coupled to the spurious impurity. Therefore at sufficiently small coupling $I\ll I^*$ ($\Delta V\le$~0) only the Kondo effect arising from the generic impurity affects the conductance. At $T\gg T_K$ the Kondo effect is not relevant and the 0.7 anomaly is restored.

Figure \ref{fig:Fig1}(b) shows the temperature dependence of the linear conductance at $\Delta V$~=~0 up to 2~K. The weak anomaly originally located at 0.75$\times 2e^2/h$ at the base temperature, decreases in conductance to about 0.6$\times 2e^2/h$ and at even higher temperature the 0.7-feature emerges as expected.
}
\section{Conclusion}
A conductance anomaly in a p-type GaAs QPC is experimentally investigated in detail. We have shown that it is possible to separate the generic 0.7-feature from spurious anomalies experimentally. By the application of different models it has been shown that anomalies in the conductance, resembling the 0.7-feature, may arise due to the presence of a chargeable impurity located nearby the QPC. However, the QPC exhibits an independent zero bias peak in the differential conductance that appears to have a different origin but is nevertheless affected when the channel is pushed towards the impurity by using asymmetric gate voltages. The splitting of this peak by in-plane and perpendicular magnetic fields suggests a Kondo-like origin for the ZBA while its zero-field splitting indicates a tunnel coupling to the impurity and relevance of two-impurity Kondo physics in this system. The results are consistent with the Kondo picture of the 0.7-anomaly and support a local origin of the ZBA. The difference to the common 0.7-anomaly is attributed to an imperfection in the channel potential.
\section{Appendix}
\begin{figure}[h!]
\centering
\includegraphics[scale=0.39]{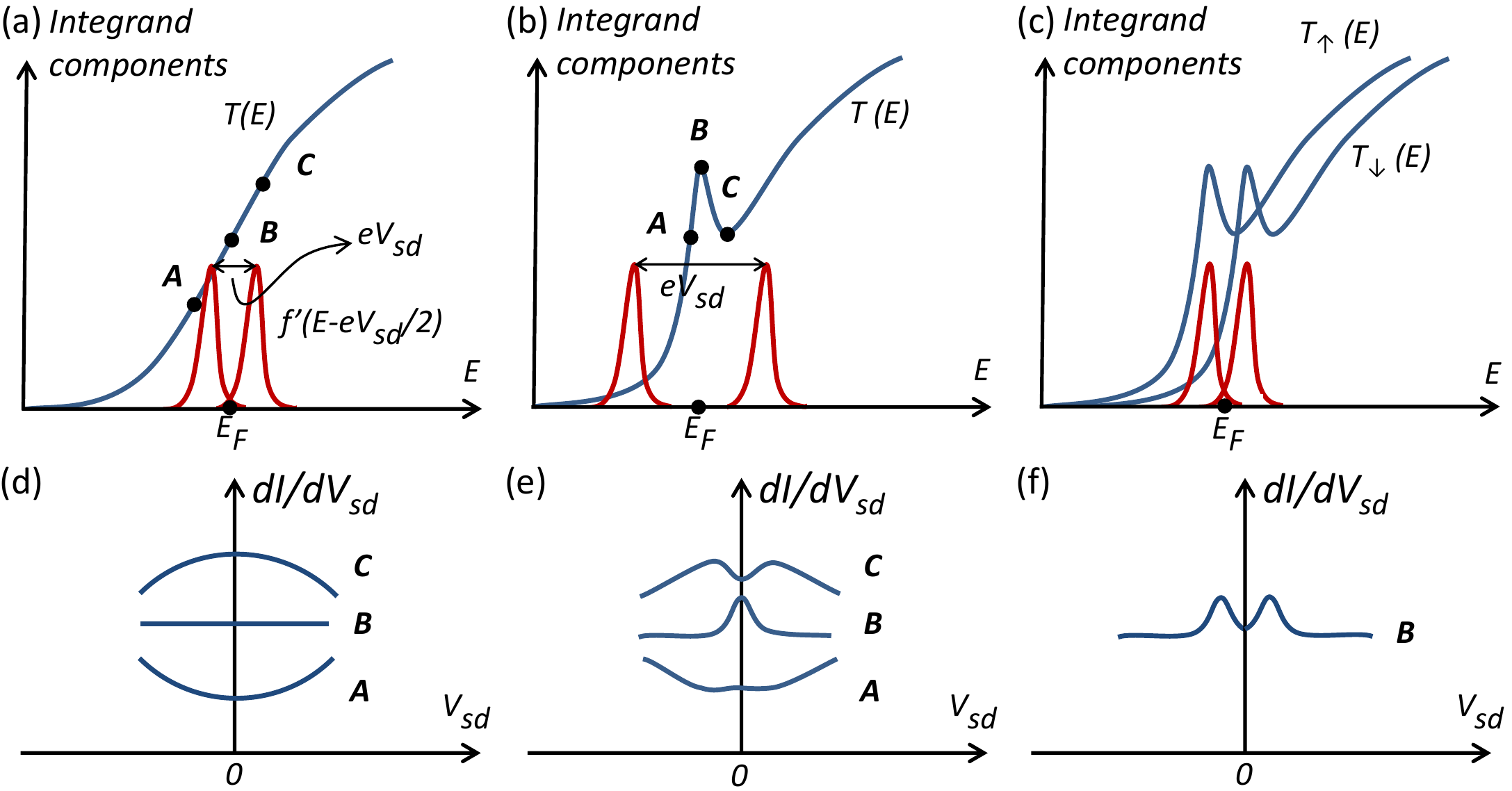}
\caption{\small Visualization of the differential conductance integral for the three cases of (a) a saddle-point potential, (b) a saddle-point potential plus a resonance and (c) a non-zero Zeeman splitting of the two subbands while the Fermi energy is at point $B$. The red curves represent the two Fermi-Dirac distribution derivatives in the integrand and the blue curve represents the transmission probability. The scales are adjusted for better visibility. (d-f) The qualitative $V_{\rm sd}$ dependence of $dI/dV_{\rm sd}$ at low biases as the Fermi energy equals to the energies labeled by $A$, $B$ and $C$ on the corresponding panels of (a-c), respectively.}\label{fig:Fig8}
\end{figure}

In this appendix we provide a qualitative explanation on how the different features observed in the linear and differential conductance may emerge from a transmission resonance. According to Landauer theory, the differential conductance of the QPC $g=dI/dV$ is calculated as \cite{ThomasIhn10}
\begin{equation}
g=-\frac{e^2}{2h}\int{dE2{\rm T}(E)\Big[f'(E+\frac{eV_{sd}}{2})+f'(E-\frac{eV_{sd}}{2})\Big]}.\nonumber
\end{equation}
The energy integral can be visualized by the schematics shown in Fig.\,\ref{fig:Fig8}. $f(E)=[1+\exp(-E/k_B{T})]^{-1}$ is the Fermi distribution (and $f'(E)$ is its derivative) with ${T}$ being the temperature. We assume that the transmission probability ${\rm T}(E)$ is independent of the bias voltage and it is merely shifted horizontally by the application of a gate voltage. The factor of $2$ in $2{\rm T}(E)$ comes from the spin degeneracy of the subband and in the presence of a Zeeman field, $2 {\rm T}(E)$ is replaced by ${\rm T}_{\uparrow}(E)+{\rm T}_{\downarrow}(E)$. The linear conductance $G$ is obtained from the zero bias ($V_{\rm sd}=0$) part of this formula. The integrand is a linear functional of the transmission ${\rm T}(E)$ and therefore different contributions to the transmission add up in the differential conductance.

First, we consider the case of a saddle-point potential whose inflection point is labelled by symbol $B$ in Fig.\,\ref{fig:Fig8}(a). If the Fermi energy is at $B$, an increasing bias does not change the differential conductance to first order and $dI/dV$ is flat at low bias voltages. At higher (lower) transmissions an increasing bias decreases (increases) the differential conductance and $dI/dV$ has a broad peak (valley) as sketched in Fig.\,\ref{fig:Fig8}(d).

Second, we consider the case of a saddle-point potential with a resonance shown in Fig.\,\ref{fig:Fig8}(b). When the Fermi energy is at the point $B$, the $dI/dV$ clearly develops a sharp peak on the top of the broad background peak or valley (caused by the saddle point potential) discussed before. On the other hand, when the Fermi energy is at point $C$, the $dI/dV$ has a minimum at zero bias, resulting in the apparent zero magnetic-field peak-splitting. The splitting of the peaks is related to both $k_BT$ and the curvature of transmission minima (Fig.\,\ref{fig:Fig8}(e)).

Lastly, we consider the case when there is a finite Zeeman splitting between the two 1D spin subbands as depicted in Fig.\,\ref{fig:Fig8}(c). The potential is assumed to be the same as in Fig.\,\ref{fig:Fig8}(b). In this case, we have two copies of all the previously-discussed features in the differential conductance plus some new features. In particular, there is a peak whenever the Fermi energy is at the resonance and the voltage bias matches the Zeeman splitting. This is visible in the sketch of $dI/dV_{\rm sd}$ at the point $B$ in Fig.\,\ref{fig:Fig8}(f). Therefore, having a peak in the differential conductance that splits with the Zeeman splitting does not unambigiously point to a Kondo effect. The ultimate proof of the Kondo effect is the observation of a splitting that is twice the Zeeman splitting.~\cite{Meir93} As we do not have an independent measure of the g-factor in our QPC, a quantitative comparison of the peak-splitting and Zeeman shift is not feasible in this case.
\begin{acknowledgments}
The authors would like to cordially thank the anonymous referee for constructive comments and suggestions which helped to improve the paper. We acknowledge stimulating discussions with I. Shurobalko and B.~R.~Bulka and thank the Swiss National Foundation for financial supports. M.~C.~is a grantee of the J\'anos Bolyai Research Scholarship of the HAS and acknowledges financial support of the European Union 7th Framework Programme (Grant No. 293797). D.~R.~and A.~D.~W.~acknowledge support from DFG SPP1285 and BMBF QuaHL-Rep 01BQ1035.
\end{acknowledgments}

%
%

\end{document}